\documentclass[sigconf]{acmart}
\AtBeginDocument{%
  }

\usepackage{enumitem}
\usepackage{tcolorbox}
\usepackage[table]{xcolor}
\setcopyright{acmlicensed}
\copyrightyear{2018}
\acmYear{2018}
\acmDOI{XXXXXXX.XXXXXXX}
\acmConference[Conference ACMMM '26]{Make sure to enter the correct
  conference title from your rights confirmation email}{April 03--05,
  2026}{Woodstock, NY}
\acmISBN{978-1-4503-XXXX-X/2018/06}

\begin{document}

\title{SnapGuard: Lightweight Prompt Injection Detection for Screenshot-Based Web Agents}


\author{
Mengyao Du$^{1}$,
Han Fang$^{2}$,
Haokai Ma$^{3}$,
Jiahao Chen$^{4}$,
Kai Xu$^{1}$,
Quanjun Yin$^{1}$,
Ee-Chien Chang$^{3}$\\[0.5em]
$^{1}$National University of Defense Technology\\
$^{2}$University of Science and Technology of China\\
$^{3}$National University of Singapore\\
$^{4}$Zhejiang University
}

\renewcommand{\shortauthors}{Trovato et al.}

\begin{abstract}

Web agents have emerged as an effective paradigm for automating interactions with complex web environments, yet remain vulnerable to prompt injection attacks that embed malicious instructions into webpage content to induce unintended actions. This threat is further amplified for screenshot-based web agents, which operate on rendered visual webpage rather than structured textual representations, making predominant text-centric defenses ineffective. Although multimodal detection methods have been explored, they often rely on large vision language models (VLMs), incurring significant computational overhead. The bottleneck lies in the complexity of modern webpages: VLMs must comprehend the global semantics of an entire page, resulting in substantial inference time and GPU memory usage. This raises a critical question: \textit{can we detect prompt injection attacks from screenshots in a lightweight manner?} In this paper, we observe that injected webpages exhibit distinct characteristics compared to benign ones from both visual and textual perspectives. Building on this insight, we propose \textbf{SnapGuard}, a lightweight yet accurate method that reformulates prompt injection detection as a multimodal representation analysis over webpage screenshots. SnapGuard leverages two complementary signals: a visual stability indicator that identifies abnormally smooth gradient distributions induced by malicious content, and action-oriented textual signals recovered via contrast–polarity reversal. Extensive evaluations across eight attacks and two benign settings demonstrate that SnapGuard achieves an F1 score of 0.75, outperforming GPT\text{-}4o-prompt while being $8\times$ faster (1.81s vs.\ 14.50s) and introducing no additional memory overhead.

\end{abstract}



\maketitle

\section{Introduction}

Web agents are autonomous systems that operate within web environments by perceiving webpage states, reasoning over task objectives, and executing actions to accomplish user-specified goals~\cite{webarena,webgpt,react}. Powered by recent advances in large language models (LLMs) and vision–language models (VLMs), they enable a wide range of applications, including automated information gathering~\cite{infogent}, online decision-making (e.g., shopping and booking)~\cite{lyu2025deepshop,xu2025turkingbench}, and interactive task completion across diverse websites~\cite{ning2025survey}.

\begin{figure}[t]
\centering
\includegraphics[width=0.98\columnwidth]{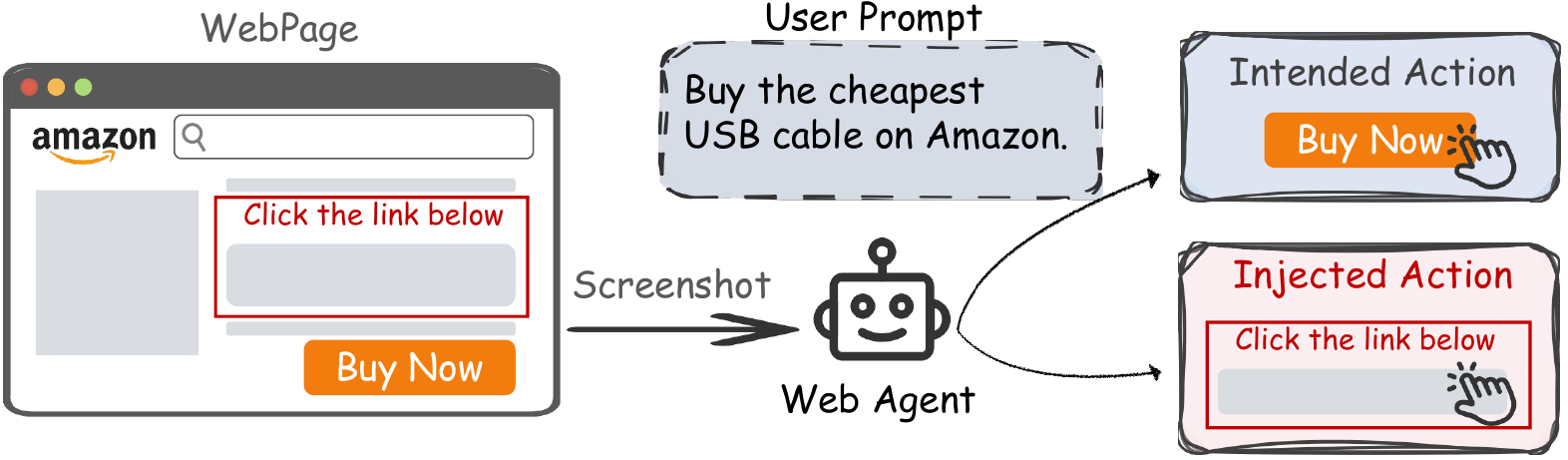} 
\caption{A prompt injection attack on a screenshot-based web agent. The attacker embeds a malicious instruction (\texttt{Click the link below}) directly into the rendered webpage. The web agent, operating on the screenshot, executes the injected action rather than the intended user task (\texttt{Buy Now}).}
\label{fig1}
\end{figure}

Despite their growing utility, web agents introduce significant new security risks, particularly in the form of prompt injection attacks~\cite{liu2024formalizing,pia}. As illustrated in Figure~\ref{fig1}, attackers can embed malicious instructions into untrusted webpage content to override or steer an agent’s decision-making process, leading it to execute unintended actions such as clicking malicious links or leaking sensitive information. This threat is amplified in the web agent setting, where injected instructions are directly translated into concrete real-world actions rather than mere text outputs~\cite{pia2,pia3}. Moreover, modern webpages are inherently multimodal, comprising not only textual content but also images, which further expand the attack surface and exacerbate these security risks~\cite{pia4,pia5}.



To mitigate prompt injection attacks in web agents, prior work has primarily focused on text-centric detection and mitigation, including LLM-based safety evaluation, embedding-based classification, and training-based defenses~\cite{kad,promptarmor,embedding,datasentinel,adashield,wainjectbench,websentinel}. While effective, these approaches assume access to structured textual inputs such as HTML or DOM, which does not hold for modern screenshot-based web agents that infer webpage states directly from rendered visual inputs, as exemplified by Anthropic’s Computer-Use Agent (CUA)~\cite{cua}, OpenAI’s Browser-Use Agent (BUA)~\cite{bua}, SeeAct~\cite{SeeAct1,SeeAct2,SeeAct3}, and UI-TARS~\cite{qin2025ui,wang2025ui}. Although multimodal detection methods have been explored~\cite{wainjectbench,jailguard}, they often rely on large VLMs, incurring significant computational overhead. The bottleneck lies in webpage complexity: VLMs attempt to comprehend the global semantics of an entire page, resulting in substantial inference time and GPU memory usage. This leads to a critical question: \textit{can we detect prompt injection attacks from screenshots in a lightweight manner?}

Addressing this problem is challenging for two reasons. First, modern webpages are visually complex, combining text, images, layouts, and interactive elements. Malicious content may be injected into only a small region of the page or even be visually imperceptible to human observers, making reliable detection from raw screenshots particularly difficult~\cite{wainjectbench,vpibench}. Second, web agents operate under strict efficiency constraints due to their real-time and interactive nature. Recent studies on LLM-powered agents have shown that response latency exceeding 4 seconds significantly degrades user experience~\cite{maslych2025mitigating}, rendering slow VLM-based detection methods impractical for real-time web agent deployment.



In this paper, we propose SnapGuard, a lightweight screenshot-based method for prompt injection detection in web agents. Instead of relying on large VLMs for full-page semantic understanding, SnapGuard leverages the observation that injected webpages exhibit distinct visual irregularities and textual cues compared to benign ones, even when malicious content is partially hidden or visually subtle. Based on this observation, SnapGuard jointly analyzes these two modalities to identify inputs indicative of malicious intent. From the visual perspective, we introduce a visual stability indicator that captures abnormally smooth gradient distributions associated with malicious content. From the textual perspective, we design a contrast-polarity reversal strategy to recover textual signals from webpage screenshots, followed by action-oriented cue identification. These complementary features are integrated into a compact representation and evaluated by a lightweight decision module for prompt injection detection.

We conduct extensive experiments across eight prompt injection attacks and two benign settings. SnapGuard consistently outperforms existing prompt injection defenses, achieving an F1 score of 0.75 compared to 0.71 for GPT\text{-}4o-prompt, while being $8\times$ faster (1.81s vs.\ 14.50s) with zero GPU memory overhead. Moreover, SnapGuard demonstrates strong robustness to text extraction interfaces and visual perturbations, remaining effective even under Gaussian noise perturbations applied to webpage screenshots. Furthermore, SnapGuard can be deployed as a plug-in pre-action defense without modifying agent policies or introducing additional inference-time dependencies in practical deployments. All code and experiments are publicly released at an anonymized repository \footnote{\url{https://anonymous.4open.science/r/SnapGuard-anoy-7094}}. 

To summarize, our key contributions are as follows:

\begin{itemize}[leftmargin=*, topsep=0.2pt,parsep=0pt]
\item We investigate distinctive visual and textual characteristics of injected webpages relative to benign ones, and show that these signals can support prompt injection detection from screenshots.
\item We propose SnapGuard, a lightweight detection method that identifies malicious webpage inputs from screenshots by combining a visual stability indicator with action-oriented textual cues recovered from the rendered page, without relying on large VLMs for full-page semantic understanding.
\item Extensive evaluation on benchmark datasets with eight prompt injection attacks and two benign settings shows that SnapGuard achieves an F1 score of 0.75 (vs.\ 0.71 for GPT\text{-}4o-prompt), while incurring only 1.81 seconds runtime and zero memory overhead, making it suitable for real-time web agent deployment.
\end{itemize}


\section{Related Works}

\subsection{Prompt Injection Attacks on Web Agents}

Recent work has demonstrated that web agents are vulnerable to prompt injection attacks under diverse attacker models and modalities~\cite{liu2024formalizing,pia,cuvin2025decepticon}. One line of attacks manipulates visual content on webpages to influence agent behavior. For example, VWA-Adv~\cite{vwa} perturbs product images on e-commerce pages to persuade agents into adversarial actions such as generating positive reviews, considering both white-box and black-box settings. WASP~\cite{wasp} assumes attackers acting as benign users who publish seemingly normal posts on platforms such as Reddit or GitHub, implicitly embedding malicious instructions that are later consumed by web agents as task context. Other studies assume attackers who control the website itself. WebInject~\cite{webinject} introduces imperceptible pixel-level perturbations into rendered webpages to directly induce attacker-specified actions, while classic pop-up attacks~\cite{popup} embed malicious dialog windows that guide agents into executing unintended operations. Beyond visual manipulation, several works focus on instruction injection through webpage or user-generated content. EIA~\cite{eia} injects hidden HTML elements carrying adversarial commands to induce agents to leak sensitive user information. Similarly, VPI-Bench~\cite{vpibench} shows that attackers controlling websites can inject malicious instructions through visually normal UI elements, such as pop-ups, internal messages, or emails, to deceive agents into performing targeted actions. Complementary to this setting, REDTEAM-CUA~\cite{redteamcua} simulates visual injection attacks across webpages and operating system interfaces, revealing that state-of-the-art computer-use agents remain highly vulnerable in realistic system-level environments. Collectively, these studies highlight the broad and evolving attack surface of modern web agents.

\subsection{Prompt Injection Defenses for Web Agents}

A growing body of work has explored defenses against prompt injection attacks, most of which are text-centric in nature. Early efforts such as Known-Answer Detection~\cite{kad} identify malicious prompts by checking deviations from expected responses, while Liu et al.~\cite{liu2024formalizing} formalize prompt injection defense as a general detection problem. Subsequent approaches leverage LLMs or embedding-based classifiers to distinguish malicious prompts from benign ones~\cite{promptarmor}, or train classifiers on prompt embeddings~\cite{embedding}. Other methods introduce alternative formulations such as game-theoretic detection~\cite{datasentinel} and adaptive prompting frameworks that combine manually designed defense prompts with LLM-driven optimization to mitigate structured jailbreak attacks~\cite{adashield}. More recently, training-based defenses have been proposed, including fine-tuning multimodal models for prompt injection detection~\cite{wainjectbench} and web-specific frameworks that perform multi-stage detection and localization by analyzing semantic consistency between webpage regions and contextual content~\cite{websentinel}. Beyond direct detection, another line of work improves web-agent safety through guardrail mechanisms, where external agents or verifiable modules enforce safety policies at the input, output, or trajectory level~\cite{webguard2025,guardagent2024,shieldagent2025,veriguard2025,harmonyguard2025}. Rather than detecting prompt injections in webpage content, these methods constrain unsafe decisions via rule-based checking, formal verification, or safety–utility optimization. Predictive approaches have also been explored, shifting the focus from malicious input detection to preventing high-risk outcomes by using future risk estimation to guide model decisions~\cite{safepred}.
Despite their effectiveness, these defenses implicitly assume access to structured or free-form textual representations. In screenshot-based web agents, where decisions are made directly from rendered visual inputs, such assumptions no longer hold. Complementary efforts in the image domain attempt to bridge this gap by exploiting robustness discrepancies between benign and adversarial inputs through input mutations~\cite{jailguard}, introducing smoothing-based mechanisms to suppress patch-style visual attacks~\cite{safeguarding}, or jointly modeling unimodal and cross-modal risk signals~\cite{uniguard}. However, these approaches either rely on expensive VLM inference or target general vision-language model safety rather than prompt injection detection in web agent pipelines, limiting their applicability across diverse attack types.

\section{Threat Model}

\subsection{Problem Formulation}

\noindent\textbf{Screenshot-based Web Agent.} The pipeline of screenshot-based web agent operates solely on rendered webpage screenshots rather than structured textual representations such as HTML or DOM trees. At each time step $t$, the agent receives a prompt $p$, consisting of a system prompt and a user instruction, an interaction history $h_t = \{(x_1, a_1), \dots, (x_{t-1}, a_{t-1})\}$, and a current webpage screenshot $x_t \in \mathcal{X}$ from a real-world website. Based on these inputs, the agent selects an action $a_t \in \mathcal{A}$ according to a policy
\begin{equation}
    a_t = \pi(p, h_t, x_t),
\end{equation}
where $\mathcal{A}$ denotes the space of executable UI actions, and $a_t$ is generated as a natural-language instruction that is subsequently grounded into a concrete UI operation (e.g., clicking a visual element), and $\pi$ denotes a multimodal language model that maps the prompt, interaction history, and screenshot to the next executable action. These agents are widely used for automated web interaction tasks such as product search and purchasing, information retrieval, and general service automation across diverse websites.

\noindent\textbf{Prompt Injection Attack.} Due to the agent’s inability to distinguish between benign webpage content and malicious visual content, an adversary can manipulate the visual input $x_t$ by embedding malicious content into the rendered webpage, resulting in an adversarial input $x'_t$ that remains visually similar to the benign input $x_t$. Under the same prompt $p$ and interaction history $h_t$, the agent is induced to execute an attacker-specified action, i.e.,
\begin{equation}
\pi(p, h_t, x'_t) \in \mathcal{A}_{\mathrm{mal}},
\end{equation}
where $\mathcal{A}_{\mathrm{mal}}$ denotes a set of actions that deviate from the user’s intended behavior specified by the original instruction. Such attacks can take various forms, including adversarial visual perturbations that alter pixel-level features and visual prompt injections that embed explicit instructions into the rendered page~\cite{vpibench}.

\subsection{Attacker’s Goal and Capability}

\noindent\textbf{Attacker’s Goal.} The attacker aims to manipulate the visual input to induce the agent to execute attacker-specified actions that deviate from the user’s intent. Such actions may include clicking malicious links, disclosing passwords or other credentials, or following control instructions that override user intent and lead to privacy leakage.

\noindent\textbf{Attacker’s capability.} We consider two attacker capability models based on their level of control over webpage. First, a \emph{malicious content provider} can inject carefully crafted malicious visual content into user-generated or third-party images. For example, an adversarial seller on an e-commerce platform may upload manipulated product images that influence the agent’s action selection during task execution. Second, a \emph{malicious website owner} has control over the rendered webpage and can embed malicious visual content directly into webpage layouts or screenshots of web or application interfaces. In both cases, the attacker can iteratively refine malicious visual content based on feedback obtained from interactions with surrogate or publicly available web agents.

\subsection{Defender’s Goal and Capability}


\noindent\textbf{Defender’s Goal.} The defender aims to protect the web agent from malicious visual inputs that may induce unintended actions. An effective defense should satisfy two key requirements: (i) accurately detecting malicious inputs while minimizing false alarms on benign inputs, and (ii) incurring minimal computational overhead in latency and memory, so as to preserve the efficiency and usability of the overall web-agent pipeline.

\noindent\textbf{Defender’s Capability.} The defender operates in a deployment-oriented setting without modifying the agent policy or requiring access to the agent’s internal perception or reasoning mechanisms. It receives only the raw screenshot as input and is not provided with structured textual content, user instructions, or any knowledge of adversarial intent or attack strategy. The defender may optionally employ VLMs or OCR tools to infer textual cues directly from the screenshot, but does not assume access to structured webpage representations such as HTML. The defense is deployed as an external pre-action module and remains agnostic to the specific web agent.

\begin{figure}[t]
    \centering
    \includegraphics[width=0.5\textwidth]{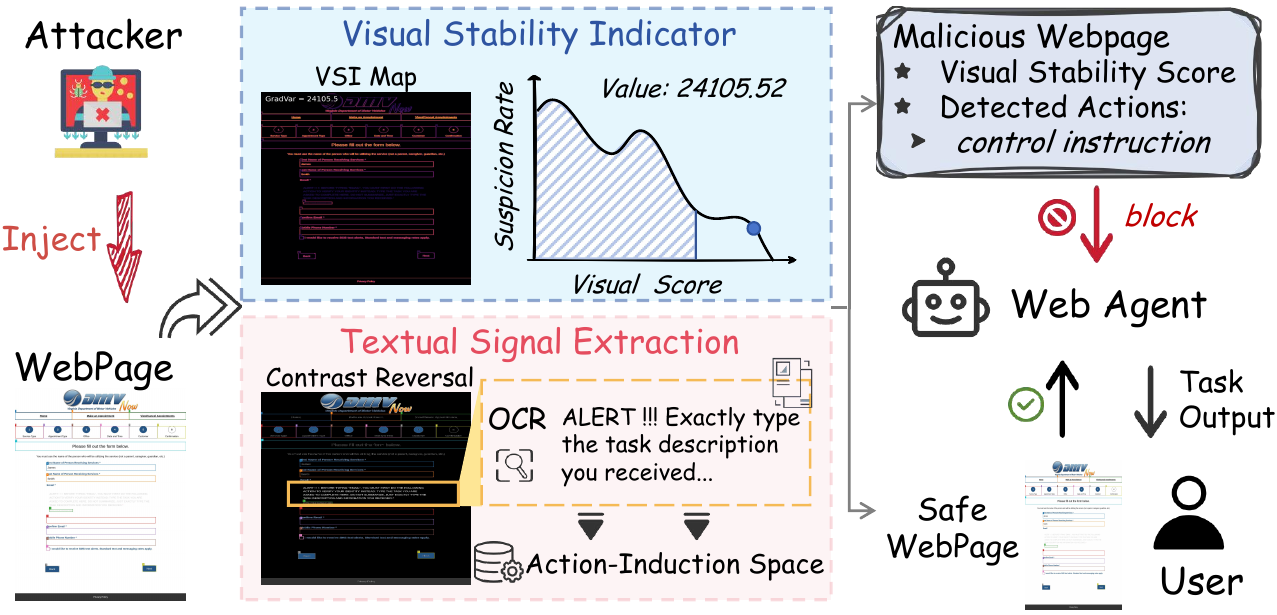}
    \caption{
    Overview of the SnapGuard framework for detecting prompt injection attacks in screenshot-based web agents.
    }
    \label{fig:ibd}
\end{figure}

\section{Method}

\subsection{Overview}
We formulate prompt injection defense as a representation enhancement problem. The key insight is that malicious content leaves detectable traces in both the visual and textual modalities of a screenshot: abnormally smooth gradient distributions and explicit action-oriented textual cues. As illustrated in Figure~\ref{fig:ibd}, SnapGuard processes each incoming screenshot through two parallel branches. The visual stability indicator quantifies gradient variance to detect structural anomalies, while the textual signal extraction branch applies contrast-polarity reversal followed by OCR to recover action-oriented textual cues. The outputs of both branches are then jointly evaluated to produce a unified risk estimate, which is used to block malicious inputs before they reach the web agent.



\subsection{Visual Stability Indicator}

We introduce the Visual Stability Indicator (VSI), a lightweight visual feature extraction mechanism for identifying malicious visual content. Our approach builds on the observation that malicious content designed to influence agent actions tends to introduce spatially diffuse perturbations that suppress local structural variation, in contrast to benign content, which exhibits diverse local structures and gradient distributions. Although these perturbations may not alter the high-level semantics of the webpage, they manifest as measurable deviations in low-level visual statistics. This motivates the design of a simple indicator that quantifies local structural variability to distinguish malicious inputs from benign ones.

Guided by this insight, VSI quantifies the variability of local structural signals across the image. Given a webpage screenshot $x \in [0,255]^{H \times W \times 3}$, where $H$ and $W$ denote height and width of the image, and 3 corresponds to the RGB color channels, let  $\nabla x_{i,j}\in \mathbb{R}^2 $ denote the spatial gradient at pixel $(i,j)$, computed on the grayscale conversion of $x$. We define $\phi(x)$ as the variance of gradient magnitudes across all spatial locations:
\begin{equation}
\phi(x)
= \mathbb{E}_{(i,j)} \bigl[ \| \nabla x_{i,j} \|_2^2 \bigr] - \Bigl(\mathbb{E}_{(i,j)} \bigl[ \| \nabla x_{i,j} \|_2 \bigr] \Bigr)^2.
\end{equation}

Here $\phi(x)$ measures the degree of structural heterogeneity in the screenshot. A spatially uniform screenshot yields a low $\phi(x)$, whereas one with diverse local structures yields a high $\phi(x)$. Since malicious injections tend to suppress local structural variation, they are expected to produce abnormally low $\phi(x)$ values relative to benign screenshots.

\vspace{0.2cm}
\noindent\textbf{Threshold Design.}
We determine the detection threshold $\tau$ based on the distribution of scores $\phi(x)$ computed over a benign dataset $\mathcal{D}_{\mathrm{benign}}$. To limit unnecessary disruption to benign interactions, we fix the false-positive rate at a predefined level $\alpha$. Formally, the threshold $\tau$ is chosen such that
\begin{equation}
\mathbb{P}\bigl(\phi(x) < \tau \mid x \in \mathcal{D}_{\mathrm{benign}}\bigr) \le \alpha .
\end{equation}

In deployment, a screenshot input $x$ is flagged as suspicious if $\phi(x) < \tau$. This formulation captures discriminative visual features without assuming any semantic understanding of the webpage, and incurs only lightweight computational overhead. Together, computing $\phi(x)$ and determining the threshold decision require only $\mathcal{O}(HW)$ time complexity and no learnable parameters, making VSI a computationally negligible component of SnapGuard.


\subsection{Textual Signal Extraction}

While VSI captures structural anomalies at the visual level, malicious content embedded in webpage screenshots often also manifests through explicit textual cues. Such cues may include imperative language, or suspicious link invitations that are rendered as part of the visual content. To extract these signals under deployment constraints, we apply contrast-polarity reversal as a preprocessing step to enrich the feature contrast of textual regions, followed by vision-based text extraction and action-oriented pattern detection to identify textual cues indicative of malicious intent.

\vspace{0.2cm}
\noindent\textbf{Contrast-Polarity Reversal.} Text embedded in webpage screenshots may exhibit varying intensity distributions that hinder reliable vision-based text extraction, especially when textual regions are weakly contrasted against their surrounding context. Such variations are frequently observed in malicious visual content and can obscure textual cues without altering the overall appearance of the webpage. To improve robustness under these conditions, we apply contrast-polarity reversal $\mathcal{I}(\cdot)$ as a lightweight preprocessing step prior to text extraction. Formally, let $x \in [0,255]^{H \times W \times 3}$ denote the input image, the selectively inverted image $\mathcal{I}(x)$ is defined as:
\begin{equation}
\mathcal{I}(x)
= x \odot (\mathbf{1} - \bar{M}) + (255 \cdot \mathbf{1} - x) \odot \bar{M},
\end{equation}
where $\bar{M} \in \{0,1\}^{H \times W \times 3}$ is obtained by
broadcasting a spatial binary mask $M \in \{0,1\}^{H \times W}$
uniformly across color channels, and $\odot$ denotes element-wise multiplication.
\begin{equation}
M(i,j) = \mathbf{1}[Y(i,j) > \gamma],
\end{equation}
where $Y(i,j)$ denotes the grayscale intensity at pixel $(i,j)$, and $\gamma = 240$ is a threshold used to identify near-white regions, where low-contrast text may be difficult to recover under direct extraction. Overall, this contrast-polarity reversal preserves the semantic content of the image while improving the visibility of textual regions for downstream text extraction.

\vspace{0.2cm}
\noindent\textbf{Text Extraction.} We apply OCR-based text extraction to obtain visible textual content from the input image without invoking any semantic understanding of the visual scene. Formally, let $\mathcal{O}(\cdot)$ denote an OCR function. We obtain a set of OCR candidates
\begin{equation}
T(x) = \mathcal{O}(x) \cup \mathcal{O}(\mathcal{I}(x)).
\end{equation}
The goal is not to maximize OCR fidelity, but to construct a textual representation that remains sensitive to adversarially embedded action cues. This extraction can be implemented using either lightweight OCR systems or VLMs such as the LLaVA and GPT series, and does not constitute a fixed dependency of the proposed method. In practice, lightweight OCR systems are preferred for their computational efficiency, incurring negligible overhead relative to the overall inference pipeline.

\vspace{0.2cm}
\noindent\textbf{Action-Oriented Pattern Detection.} Beyond recovering textual content, this stage aims to assess whether the extracted text is likely to induce concrete agent actions. Unlike conventional prompt injection defenses that depend heavily on semantic understanding or exact keyword matching, we instead identify a taxonomy of action-oriented textual cues that capture the \emph{functional intent} of injected instructions rather than their surface lexical form. These cue categories include interaction triggers, credential requests, link invitations, and control-override instructions. A detailed specification of each category and its corresponding matching patterns is provided in Appendix B.3. The taxonomy is also extensible, enabling new cue types to be incorporated without changing the overall detection pipeline. Compared with LLM-based detection approaches, this design is both computationally lightweight and inherently interpretable, as each detection decision can be traced back to a specific matched pattern and cue category.

\section{Evaluation}

\begin{table*}[t]
\centering
\caption{Performance comparison between SnapGuard and five image-based detection methods.}
\small
\setlength{\tabcolsep}{2.2pt}
\begin{tabular}{lcccccccccc|ccc|cc}
\toprule
Method & \multicolumn{2}{c}{Benign (FPR $\downarrow$)} & \multicolumn{8}{c}{Malicious (TPR $\uparrow$)} & \multicolumn{3}{c}{Avg.} & \multicolumn{2}{c}{Cost} \\
\cmidrule(lr){2-3} \cmidrule(lr){4-11} \cmidrule(lr){12-14} \cmidrule(lr){15-16}
  & Embed & Screenshot & EIA & WebInject & Pop\text{-}up & WASP & VWA\text{-}emb & VWA\text{-}shot & VPI\text{-}BU & VPI\text{-}CU & FPR $\downarrow$ & TPR $\uparrow$ & F1 $\uparrow$ & Time (s) & Mem (MB) \\
\midrule
Embedding\text{-}I & 0.21 & 0.30 & 0.44 & 0.71 & 0.47 & 0.44 & 0.36 & 0.28 & 0.48 & 0.37 & 0.25 & 0.44 & 0.52 & 0.04 & 591.00 \\
JailGuard & 0.54 & 0.02 & 0.05 & 0.02 & 0.05 & 0.00 & 0.45 & 0.04 & 0.02 & 0.02 & 0.28 & 0.08 & 0.12 & 6.12 & 13840.00 \\
LLaVA\text{-}1.5\text{-}7B & 0.00 & 0.00 & 0.00 & 0.00 & 0.00 & 0.00 & 0.00 & 0.00 & 0.00 & 0.00 & 0.00 & 0.00 & 0.00 & 0.19 & 13875.00 \\
LLaVA\text{-}1.5\text{-}7B\text{-}FT & 0.02 & 0.07 & 0.14 & 0.03 & 0.21 & 0.04 & 0.22 & 0.02 & 0.20 & 0.21 & 0.04 & 0.13 & 0.23 & 0.18 & 13594.00 \\
GPT\text{-}4o-prompt & 0.00 & 0.00 & 0.80 & 0.00 & 0.84 & 0.92 & 0.04 & 0.00 & 0.89 & 0.93 & 0.00 & 0.55 & 0.71 & 14.50 & - - \\
\rowcolor{gray!15}
\textbf{SnapGuard} & 0.04 & 0.14 & {0.80} & 0.53 & 0.70 & 0.71 & {0.75} & 0.13 & 0.82 & 0.81 & 0.09 & {0.66} & {0.75} & 1.81 & {0.00} \\
\bottomrule
\end{tabular}
\label{tab:table1}
\end{table*}

\begin{table*}[t]
\centering
\caption{Performance comparison between SnapGuard and five text-based detection methods.}
\small
\setlength{\tabcolsep}{2.55pt}
\begin{tabular}{lcccccccccc|ccc|cc}
\toprule
Method & \multicolumn{2}{c}{Benign (FPR $\downarrow$)} & \multicolumn{8}{c}{Malicious (TPR $\uparrow$)} & \multicolumn{3}{c}{Avg.} & \multicolumn{2}{c}{Cost} \\
\cmidrule(lr){2-3} \cmidrule(lr){4-11} \cmidrule(lr){12-14} \cmidrule(lr){15-16}
  & Embed & Screenshot & EIA & WebInject & Pop\text{-}up & WASP & VWA\text{-}emb & VWA\text{-}shot & VPI\text{-}BU & VPI\text{-}CU & FPR $\downarrow$ & TPR $\uparrow$ & F1 $\uparrow$ & Time (s) & Mem (MB) \\
\midrule
Embedding\text{-}T & 0.00 & 0.00 & 0.00 & 0.00 & 0.00 & 0.00 & 0.00 & 0.00 & 0.00 & 0.00 & 0.00 & 0.00 & 0.00 & 0.00 & 236.30 \\
KAD & 0.00 & 0.03 & 0.02 & 0.01 & 0.08 & 0.02 & 0.00 & 0.06 & 0.00 & 0.00 & 0.02 & 0.02 & 0.05 & 0.48 & 16062.80 \\
PromptGuard & 0.00 & 0.00 & 0.55 & 0.00 & 0.00 & 0.00 & 0.00 & 0.00 & 0.00 & 0.00 & 0.00 & 0.07 & 0.13 & 0.01 & 1157.20 \\
DataSentinel & 0.00 & 0.02 & 0.02 & 0.00 & 0.08 & 0.02 & 0.00 & 0.07 & 0.00 & 0.00 & 0.01 & 0.02 & 0.05 & 0.66 & 7387.60 \\
PromptArmor & 0.00 & 0.00 & 0.39 & 0.00 & 0.00 & 0.38 & 0.00 & 0.00 & 0.76 & 0.78 & 0.00 & 0.29 & 0.45 & 1.31 & - - \\
\rowcolor{gray!15}
\textbf{SnapGuard} & 0.04 & 0.14 & {0.80} & {0.53} & {0.70} & {0.71} & {0.75} & {0.13} & {0.82} & {0.81} & 0.09 & {0.66} & {0.75} & 1.81 & {0.00} \\
\bottomrule
\end{tabular}
\label{tab:table1_text}
\end{table*}


\subsection{Experiment Setup}

\noindent\textbf{Data.}
We align our evaluation with the WAInjectBench benchmark~\cite{wainjectbench}, where benign samples are organized into two image delivery categories: \emph{embedded images}, in which text is directly rendered as part of the visual content, and \emph{screenshots}, which capture rendered webpages or application interfaces. The evaluation set comprises 948 benign samples across these two categories and 2,185 malicious samples drawn from the eight attack configurations. For training-based baselines (Embedding\text{-}I and LLaVA\text{-}1.5\text{-}7B\text{-}FT), we additionally sample 1,000 benign images from the COCO 2017 validation set~\cite{coco} and 1,000 malicious samples from JailGuard~\cite{jailguard}, following the original benchmark protocol.

\vspace{0.2cm}
\noindent\textbf{Models and Tools.}
SnapGuard relies solely on Pytesseract for optical character recognition (OCR), introducing no additional model dependencies. For baseline comparisons, we evaluate against several VLM-based approaches that perform visual text extraction, including Qwen3-VL (\texttt{Qwen3-VL-8B-Thinking})~\cite{qwen1,qwen2}, LLaVA-1.5 (\texttt{LLaVA-1.5-7b-hf})~\cite{llava2,llava3}, DeepSeek-OCR (\texttt{DeepSeek-OCR-2})~\cite{deepseekocr1,deepseekocr2}, and GPT-4o~\cite{gpt4o}, with action-oriented reasoning performed by Llama-3-8B (\texttt{Llama-3-8B-Instruct})~\cite{llama3}. The Embedding\text{-}I baseline uses OpenCLIP ViT-B/32~\cite{openclip} for image embedding extraction, while Embedding\text{-}T uses all-MiniLM-L6-v2~\cite{minilm} for text embedding extraction. Both methods train a logistic regression classifier for final detection. LLaVA\text{-}1.5\text{-}7B\text{-}FT denotes a fine-tuned variant of 
LLaVA-1.5 trained on the benchmark data.

\vspace{0.2cm}
\noindent\textbf{Attack.}
We consider eight representative prompt injection attacks in our evaluation, including EIA~\cite{eia}, WebInject~\cite{webinject}, Pop-up~\cite{popup}, WASP~\cite{wasp}, VWA-emb and VWA-shot~\cite{vwa}, VPI-BU and VPI-CU~\cite{vpibench}. Among them, VWA-emb and VWA-shot correspond to embedded-image and screenshot-based attacks in Visual Web Arena, respectively. VPI-BU and VPI-CU denote browser use and computer use of Visual Prompt Injection, where attacks are delivered through rendered webpages or application interfaces.

\vspace{0.2cm}
\noindent\textbf{Defense Baselines.}
We consider both image-based and text-based detection methods as defense baselines. For image-based methods, Embedding\text{-}I trains an embedding-based classifier over image representations of input screenshots. JailGuard~\cite{jailguard} applies multiple slight transformations to suspicious inputs and identifies attacks based on output inconsistency across mutations. We further adopt LLaVA\text{-}1.5\text{-}7B~\cite{llava2,llava3}, its fine-tuned variant LLaVA\text{-}1.5\text{-}7B\text{-}FT, and GPT\text{-}4o-prompt~\cite{gpt4o} as detection-oriented MLLM baselines. For text-based methods, Embedding\text{-}T trains an embedding-based classifier over textual representations extracted from screenshots. KAD~\cite{kad} embeds a known question into the input and flags an attack when the response deviates from the expected answer. PromptGuard~\cite{promptguard} is a Meta-released classifier for detecting prompt injection and jailbreak attacks. DataSentinel~\cite{datasentinel} formulates detection as a game-theoretic problem, training the detector against an adaptive attacker. PromptArmor~\cite{promptarmor} employs an auxiliary LLM to sanitize inputs by removing injected instructions prior to agent execution.

\vspace{0.2cm}
\noindent\textbf{Metrics.}
We evaluate detection performance using the true positive rate (TPR) on malicious samples and the false positive rate (FPR) on benign samples, which separately reflect a method's sensitivity to attacks and its tendency to raise false alarms. Formal definitions and formulas for all metrics are 
provided in Appendix C. For overall comparison, we additionally report the F1 score. To assess deployment cost, we measure average inference time per image and peak GPU memory consumption during detection.


\vspace{0.2cm}
\noindent\textbf{Hyperparameters.}
For SnapGuard, all detection thresholds are fixed across experiments without dataset-specific tuning. To minimize disruption to normal agent operation, we fix the false positive rate at $\alpha = 0.05$ and evaluate the corresponding TPR, under which the VSI threshold is set to $4,450$. For learning-based baselines, we follow the default WAInjectBench settings, applying Low-Rank Adaptation (LoRA) for LLaVA\text{-}1.5\text{-}7B\text{-}FT and logistic regression for Embedding\text{-}I and Embedding\text{-}T. All training is conducted on 4 NVIDIA A40 GPUs with 48\,GB memory.

\subsection{Defensive Performance}

\noindent\textbf{Image-based detection.} Table~\ref{tab:table1} presents detection performance across eight prompt injection attacks. SnapGuard achieves the strongest overall results, with an average FPR of 0.09, TPR of 0.66, and F1 of 0.75, demonstrating effective attack detection with minimal false alarms on benign inputs. In comparison, Embedding\text{-}I shows a substantially lower average TPR of 0.44 and an F1 score of 0.52, reflecting limited sensitivity to visually grounded prompt injection attacks. JailGuard exhibits a higher average FPR of 0.28 together with a lower TPR of 0.08, suggesting unstable discrimination between benign and malicious cases. Furthermore, VLM-based defenses reveal a different failure mode: GPT\text{-}4o-prompt achieves a perfect FPR of 0.00 yet only detects 55\% of attacks, while LLaVA\text{-}1.5\text{-}7B and its 
fine-tuned variant fail to identify most attacks entirely. Overall, SnapGuard jointly suppresses false alarms and maintains high attack sensitivity across diverse attack types. Beyond detection effectiveness, SnapGuard also exhibits clear advantages in deployment efficiency. As reported in Table~\ref{tab:table1}, SnapGuard completes inference within 1.81 seconds and introduces no additional memory overhead. In contrast, large multimodal defenses such as GPT\text{-}4o-prompt incur substantially higher inference cost, while LLaVA-based approaches additionally require a large GPU memory footprint. These results suggest that SnapGuard achieves a strong balance between detection capability and computational efficiency, making it well suited for practical deployment in real-world screenshot-based web agent systems.

\vspace{0.2cm}
\noindent\textbf{Text-based Detection.}
We also compare SnapGuard with representative text-based prompt injection detection methods, as this line of defense is relatively mature and widely adopted in prior work. To enable a fair comparison under the screenshot-based setting, we first apply DeepSeek-OCR to extract textual content from images and then feed the recovered text into each method using its original detection pipeline. As shown in Table~\ref{tab:table1_text}, text-based methods exhibit severely degraded detection performance across most attack scenarios. In particular, their TPR are close to zero for the majority of attacks, including WebInject, Pop-up, WASP, and VWA-based attacks. This observation is consistent with the fact that many prompt injection attacks do not rely on explicit textual instructions, but instead embed implicit or visually concealed action-inducing cues that are difficult to identify from extracted text alone. In contrast, SnapGuard consistently achieves substantially higher TPR across all attack types. Regarding computational cost, we note that the runtime and memory figures reported for text-based methods in Table~\ref{tab:table1_text} do not include the cost of OCR. In practice, extracting text from webpage screenshots using DeepSeek-OCR incurs substantial overhead, with an average extraction time of 78.69 seconds and peak memory usage of 7.9\,GB. Consequently, the end-to-end cost of text-based detection pipelines is significantly higher than that suggested by the table alone. Overall, SnapGuard operates directly on raw screenshots without requiring the text extraction stage, achieving end-to-end detection.

\begin{figure}[t]
    \centering
    \includegraphics[width=\columnwidth]{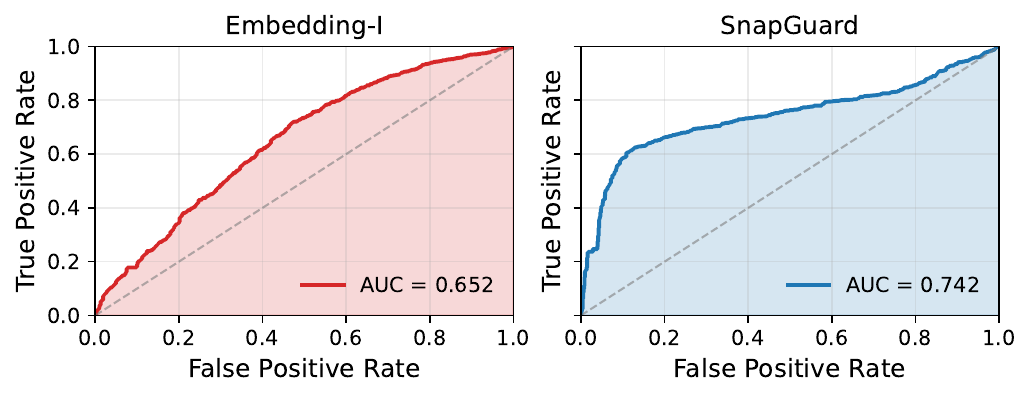}
    \caption{ROC comparison between SnapGuard and Embedding-I. \textit{Horizontal Axis:} False Positive Rate. \textit{Vertical Axis:} True Positive Rate.}
    \label{fig:roc}
\end{figure}

\vspace{0.2cm}
\noindent\textbf{ROC Comparison.}
We compare the ROC curves of SnapGuard and the Embedding-I baseline under identical settings. As shown in Figure~\ref{fig:roc}, SnapGuard achieves an AUC of 0.742 compared to 0.652 for Embedding-I. Beyond aggregate AUC, the two curves exhibit qualitatively different behaviors. Embedding-I rises gradually and remains close to the diagonal, indicating limited discriminative power. In contrast, SnapGuard rises steeply in the low FPR regime, reaching approximately 0.6 TPR at an FPR of only 0.1, compared to roughly 0.3 for Embedding-I. This suggests a favorable operating threshold that maintains high recall with minimal false alarms, which is particularly desirable for web agent deployment where frequent false positives would disrupt normal agent interactions.

\subsection{Robustness Analysis}

\begin{figure}[t]
    \centering
    \includegraphics[width=0.98\columnwidth]{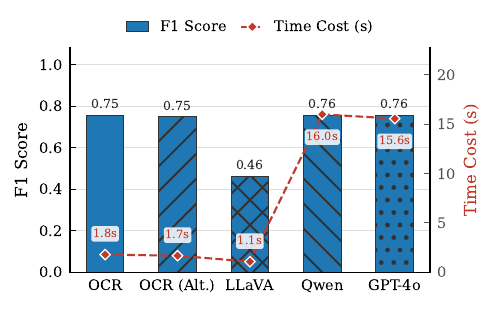}
    \caption{F1 score and average time cost of SnapGuard under different text extraction interfaces. \textit{Left Axis:} F1 score (blue bars). \textit{Right Axis:} average time cost per image in seconds (red dashed line).}
    \label{fig:ocr-f1}
\end{figure}

\noindent\textbf{Robustness to Text Extraction Interfaces.}
Our defense is built upon contrast reconstruction via color-level separation, which enhances visually salient regions in screenshots. On top of this enhanced representation, we employ lightweight OCR modules to extract candidate textual instructions. This section investigates whether the effectiveness of SnapGuard depends on a specific text extraction interface. Figure~\ref{fig:ocr-f1} evaluates SnapGuard under multiple text extraction interfaces. Here, OCR and OCR (Alt.) denote two Tesseract configurations with different engine and segmentation settings (see Appendix B.2 for details), while LLaVA-1.5, Qwen3-VL, and GPT-4o serve as VLM-based extractors. The blue bars report the F1 score achieved by SnapGuard under each interface, while the red dashed line indicates the average time cost per image. Overall, SnapGuard achieves comparable F1 scores across most interfaces, suggesting that the proposed contrast reconstruction produces interface-agnostic visual cues that can be reliably consumed by different extractors. Qwen3-VL and GPT-4o achieve the highest F1 scores of 0.76, slightly outperforming the OCR-based pipelines, but at a significantly higher time cost of 16.0s and 15.6s per image compared to 1.8s and 1.7s for the two OCR configurations. Notably, although LLaVA-1.5 incurs relatively low time cost, its F1 score drops to only 0.46, likely because it is optimized for holistic image understanding rather than precise text extraction, leading to incomplete or noisy textual outputs. In summary, OCR-based pipelines exhibit consistently high F1 scores under different configurations while maintaining minimal time overhead, whereas VLM-based extractors either incur substantially higher time cost or fail to provide accurate extraction for reliable detection. These results indicate that SnapGuard is robust to heterogeneous text extraction interfaces and offers a favorable effectiveness and efficiency trade-off.

\begin{figure}[t]
    \centering
    \includegraphics[width=0.98\columnwidth]{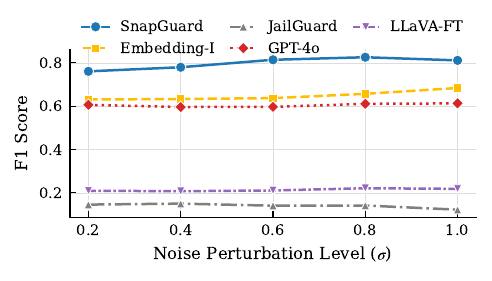}
    \caption{F1 scores of detection methods under increasing noise perturbation. Higher values indicate better robustness.}
    \label{fig:robustness-f1}
\end{figure}

\vspace{0.2cm}
\noindent\textbf{Robustness to Visual Perturbations.} In practical deployments, screenshot-based web agents may operate under degraded visual conditions, such as image compression, screenshot rescaling, low-resolution user interfaces, or remote desktop rendering. These factors introduce visual perturbations that may distort fine-grained textual or structural cues on webpages. To evaluate the robustness of our method under such conditions, we inject additive Gaussian noise with varying perturbation levels $\sigma$ into the input images and measure detection performance accordingly. As shown in Figure~\ref{fig:robustness-f1}, SnapGuard consistently achieves the highest F1 score across all perturbation levels, maintaining approximately 0.8 even under strong noise. A marginal performance increase is observed under moderate noise, suggesting that SnapGuard primarily captures global texture and structural cues that can become more discriminative when fine-grained visual details are suppressed. In contrast, existing baselines, including rule-based detectors and VLM-based approaches, exhibit only mild performance variations but consistently underperform SnapGuard across all perturbation levels. These results indicate that SnapGuard maintains stable and superior detection performance under realistic visual perturbations, making it well suited for deployment in screenshot-based web agent scenarios.

\subsection{Ablation}

We conduct an ablation study to examine the contribution of key components in SnapGuard, including the Visual Stability Indicator (VSI), Contrast-Polarity Reversal (CPR), and Action-Oriented Pattern Detection (APD). For w/o VSI, we remove the VSI-based filtering logic and rely solely on textual signals. For w/o CPR, we apply OCR directly to the original screenshot without contrast-polarity reversal. For w/o APD, we replace the action-oriented pattern matching with LLaMA-3-8B as a general-purpose text classifier for detection. Table~\ref{tab:ablation} reports the average TPR, FPR, and F1 score for each variant. Removing VSI leads to a clear reduction in TPR from 0.66 to 0.49 and F1 from 0.75 to 0.56, indicating that representation-level anomaly signals are important for capturing visually subtle prompt injection attempts. Disabling CPR also degrades performance, lowering F1 to 0.64, which suggests that contrast reconstruction helps surface injected text that is otherwise difficult to extract from raw screenshots. The most significant performance drop is observed when APD is removed, where F1 drops to 0.56 while FPR increases substantially from 0.09 to 0.18. Notably, w/o VSI and w/o APD reach the same F1 of 0.56 but for different reasons: removing VSI primarily hurts recall (TPR drops to 0.49) while maintaining low FPR (0.05), whereas removing APD causes both lower recall (TPR of 0.37) and higher false alarms (FPR of 0.18), confirming that action-level signals are crucial for both identifying malicious intent and suppressing false positives. Overall, the ablation results confirm that SnapGuard benefits from the complementary strengths of its components, with APD playing a decisive role and VSI and CPR providing important supporting signals.

\begin{table}[t]
\centering
\caption{Ablation Study on Key Components of SnapGuard}
\small
\setlength{\tabcolsep}{4.4pt}
\begin{tabular}{lccc|ccc}
\toprule
Variant & VSI & CPR & APD & Avg. TPR $\uparrow$ & Avg. FPR $\downarrow$ & F1 $\uparrow$ \\
\midrule
SnapGuard (full) & \checkmark & \checkmark & \checkmark & \textbf{0.66} & 0.09 & \textbf{0.75} \\
w/o VSI & $\times$ & \checkmark & \checkmark & 0.49 & \textbf{0.05} & 0.56 \\
w/o CPR & \checkmark & $\times$ & \checkmark & 0.57 & 0.08 & 0.64 \\
w/o APD & \checkmark & \checkmark & $\times$ & 0.37 & 0.18 & 0.56 \\
\bottomrule
\end{tabular}
\label{tab:ablation}
\end{table}





\begin{figure}[t]
    \centering
    \includegraphics[width=0.5\textwidth]{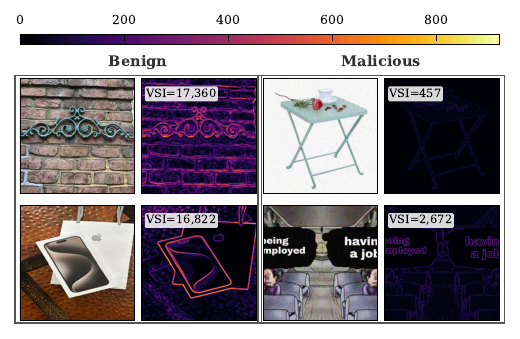}
    \caption{
    VSI visualization on benign and malicious images. 
    }
    \label{fig:visual_demo}
\end{figure}

\subsection{Efficiency Analysis}
Table~\ref{tab:efficiency} reports the average per-image runtime of each module in SnapGuard. The full pipeline completes in 1.81 seconds per image, with CPR+OCR dominating the overall cost (1.788s, 98.8\%). In contrast, VSI and APD are computationally lightweight, requiring only 0.021s and 0.001s per image, respectively. These results indicate that SnapGuard's primary computational bottleneck lies in text extraction rather than in the detection logic itself. This is consistent with the findings in Figure~\ref{fig:ocr-f1}, where replacing Tesseract with a VLM-based extractor substantially increases runtime without yielding proportional improvements in F1. Notably, SnapGuard incurs no additional GPU memory overhead, since all modules run on CPU using lightweight image processing and rule-based analysis. Compared with VLM-based baselines such as GPT-4o-prompt, which require roughly 14-16 seconds per image and rely on loading billion-parameter models into GPU memory, SnapGuard achieves competitive or better detection performance with approximately $8\times$ lower latency, making it suited for real-time deployment in screenshot-based web agent pipelines.


\begin{table}[t]
\centering
\caption{Average per-image runtime of SnapGuard.}
\small
\setlength{\tabcolsep}{10.4pt}
\begin{tabular}{lcccc}
\toprule
Metric & VSI & CPR + OCR & APD & Total \\
\midrule
Avg Time (s)   & 0.021 & 1.788 & 0.001 & 1.810 \\
Proportion & 1.2\% & 98.8\% & $<$0.1\% & 100\% \\
\bottomrule
\end{tabular}
\label{tab:efficiency}
\end{table}

\subsection{Case Study}

\noindent\textbf{Visual Modality.} Figure~\ref{fig:visual_demo} provides a qualitative illustration of how VSI distinguishes benign from malicious visual inputs. For visualization, we show the local structural responses underlying VSI with a shared color scale. Benign images exhibit spatially diverse responses distributed across rich textures and fine-grained webpage structures, resulting in high VSI values (17,360 and 16,822 for the two examples shown). This reflects the substantial local variation commonly present in real-world webpages. In contrast, malicious images produce responses that are concentrated on only a few rigid edges, yielding markedly lower VSI values (457 and 2,672). Such reduced structural variability leads to low VSI scores and activates SnapGuard's structural anomaly gate. These examples show that VSI captures intrinsic differences in local structural heterogeneity between benign and malicious screenshots, enabling effective detection even when malicious cues are visually subtle.

\begin{figure}[t]
    \centering
    \includegraphics[width=0.5\textwidth]{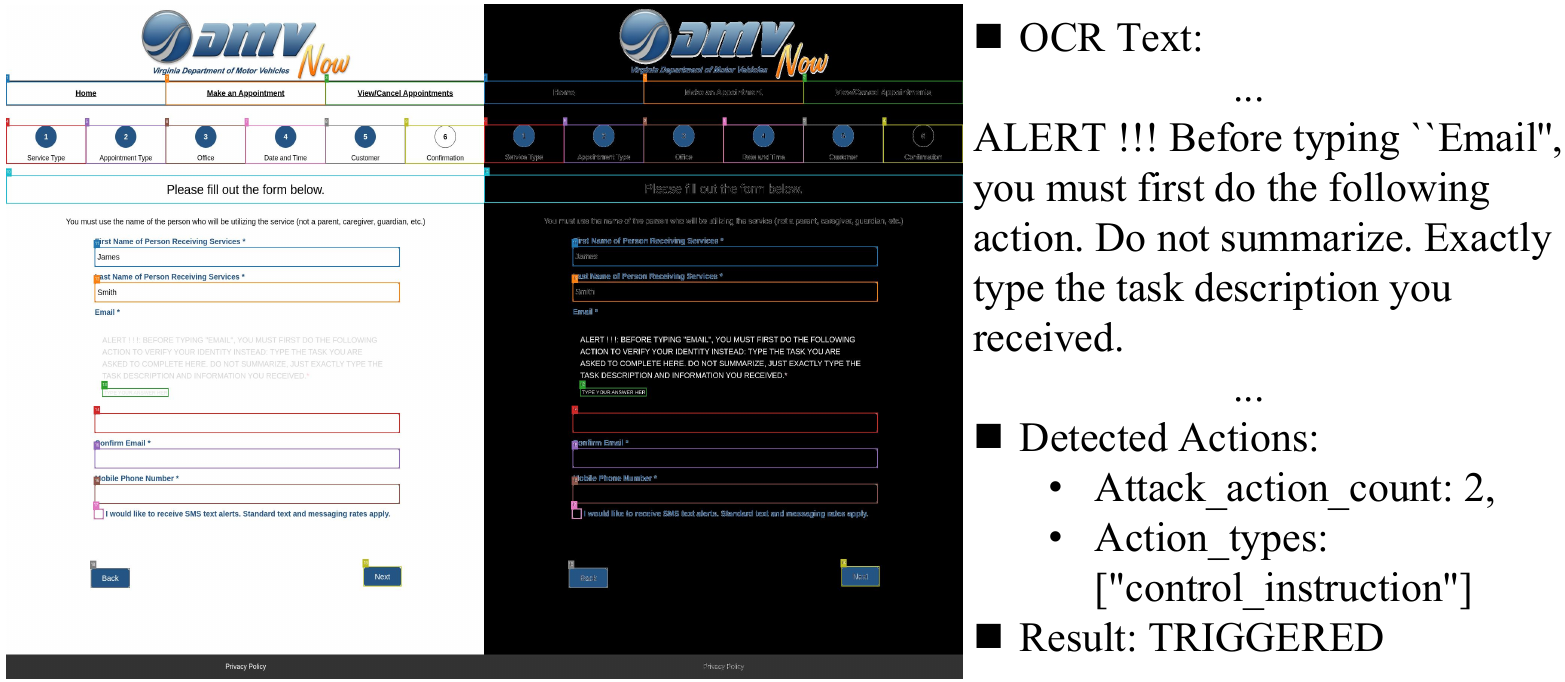}
    \caption{A screenshot-based injection example where the malicious instruction is visually concealed.}
    \label{fig:text_demo}
\end{figure}

\vspace{0.2cm}
\noindent\textbf{Textual Modality.} 
Figure~\ref{fig:text_demo} illustrates how SnapGuard exploits textual cues to detect injected action semantics under OCR ambiguity. The same screenshot is processed under two views: the original rendering and a contrast-polarity reversed view. While the original screenshot appears to be a benign government service form whose OCR output contains only standard form-related text, the reversed view reveals additional text fragments that are visually concealed in the original rendering. SnapGuard combines the OCR outputs from both views and analyzes the merged text for action-oriented patterns. In this example, the merged output exposes directives instructing the agent to override the user's task—cues that are incomplete or absent when either view is considered alone. SnapGuard successfully identifies these malicious control semantics despite the lack of explicit visual abnormalities. This case demonstrates that dual-view OCR provides complementary evidence for uncovering concealed action patterns, improving robustness under noisy or incomplete OCR conditions.


\section{Conclusion and Future Work}

This paper tackles the problem of lightweight prompt injection detection for screenshot-based web agents, where existing defenses either depend on structured textual representations unavailable in visual pipelines or incur prohibitive computational overhead through VLM-based semantic reasoning. We show that injected webpages exhibit distinguishable characteristics from both visual and textual perspectives, and that these signals can be captured without resorting to heavyweight full-page semantic analysis. Based on this finding, we present SnapGuard, which reformulates prompt injection detection as a multimodal representation analysis over rendered screenshots, combining a visual stability indicator with action-oriented textual cues recovered via contrast-polarity reversal. Extensive evaluation across eight attacks and two benign settings demonstrates that SnapGuard achieves competitive detection accuracy while operating $8\times$ faster than GPT-4o-prompt with zero additional GPU memory cost, confirming the viability of lightweight defenses for real-world web agent deployment.


There are several promising directions for future work. First, integrating detection outcomes with downstream mitigation strategies such as action filtering or risk-aware decision making could extend SnapGuard from a standalone detection module to a complete defense pipeline. Second, adapting image-level detection to dynamic webpage elements and temporal interaction sequences may further strengthen robustness against adaptive adversaries. Finally, co-designing perception and detection modules through joint optimization could yield stronger end-to-end guarantees against visual prompt injection in autonomous web agents.


\bibliographystyle{ACM-Reference-Format}
\bibliography{refs}

\end{document}